\documentclass[twocolumn,bbm,twocolappendix]{aastex631}

\usepackage{graphicx}
\usepackage{natbib}
\usepackage{amsmath}
\usepackage{mathtools}
\usepackage{amssymb}
\usepackage{wasysym}
\usepackage{verbatim}
\usepackage{txfonts}
\usepackage[T1]{fontenc}
\usepackage{ae, aecompl}
\usepackage{hyperref}
\usepackage{upgreek}
\usepackage{xcolor}
\usepackage{latexsym,longtable,epsf,ulem}
\usepackage{cases} 
\usepackage{enumerate} 
\usepackage{booktabs}
\newcommand{\EQ}{\begin{equation}}
\newcommand{\EN}{\end{equation}}
\newcommand{\EQA}{\begin{eqnarray}}
\newcommand{\ENA}{\end{eqnarray}}

\newcommand{\kms}{\,{\rm km\,s^{-1}}}

\newcommand{\cmn}{\,{\rm cm^{-3}}}
\newcommand{\yr}{\,{\rm yr}}

\newcommand{\ted}{{t_{\rm ed}}}

\newcommand{\lf}{{\ell_{\rm f}}}
\newcommand{\kf}{{k_{\rm f}}}
\newcommand{\bfl}{\mbox{\boldmath $l$} {}}
\newcommand{\urms}{{u_{\rm rms}}}
\newcommand{\mrms}{\mathcal{M}_{\rm rms}}
\newcommand{\brms}{{B_{\rm rms}}}

\newcommand{\kk}{\mbox{\boldmath $k$} {}}
\newcommand{\dd}{{\rm d} {}}
\newcommand{\UU}{\mbox{\boldmath $U$} {}}
\newcommand{\BB}{\mbox{\boldmath $B$} {}}
\newcommand{\FF}{\mbox{\boldmath $F$} {}}
\newcommand{\JJ}{\mbox{\boldmath $J$} {}}
\newcommand{\SSS}{\mbox{\boldmath $S$} {}}
\newcommand{\mkG}{\,{\rm \upmu G}}

\newcommand{\B}{{\bm B}}
\newcommand{\sigmarm}{\bar{\sigma}_{\rm RM}}

\newcommand{\edense}{\bar{n}_{\rm e}}
\newcommand{\mbmag}{\langle B\rangle}
\newcommand{\mrho}{{\langle \rho\rangle}}

\newcommand{\kpc}{\, {\rm kpc}}
\newcommand{\pc}{\, {\rm pc}}
\newcommand{\radm}{\,{\rm rad\,m^{-2}}}

\def\Rm{{\rm Rm}}
\def\Rmc{{\rm Rm}_{\rm cr}}
\def\Rey{{\rm Re}}
\def\Pm{\rm Pm}

\shorttitle{Fluctuation dynamos in supersonic turbulence}
\shortauthors{Nagdeo, Sur \& Vaidya}

\begin{document}

\title{Fluctuation dynamos in supersonic turbulence at $\Pm \gtrsim 1$}

\correspondingauthor{Ameya Uday Nagdeo}
\email{ameya.nagdeo@iiap.res.in}

\author[0009-0001-8508-3487]{Ameya Uday Nagdeo }
\affiliation{Indian Institute of Astrophysics, 2nd Block, Koramangala, Bangalore 560034, India}
\affiliation{Affiliated to Pondicherry University, R.V. Nagar, Kalapet, Puducherry, 605014, India}

\author[0000-0003-4286-8476]{Sharanya Sur}
\affiliation{Indian Institute of Astrophysics, 2nd Block, Koramangala, Bangalore 560034, India}
\affiliation{Affiliated to Pondicherry University, R.V. Nagar, Kalapet, Puducherry, 605014, India}

\author[0000-0001-5424-0059]{Bhargav Vaidya}
\affiliation{Indian Institute of Technology-Indore, Simrol, 452020, India}

\begin{abstract}

Fluctuation dynamos provide a robust mechanism for amplifying weak seed magnetic fields 
in turbulent astrophysical plasmas. However, their behaviour in the highly compressible 
regimes characteristic of the interstellar medium remains incompletely understood. 
Using high-resolution 3D magnetohydrodynamic simulations of supersonic turbulence with 
rms Mach number $\mrms \approx 11$, we explore fluctuation dynamos across magnetic 
Prandtl numbers $\Pm = 1$--10. At $\Pm=1$, dynamo growth is slower and saturates at 
lower magnetic-to-kinetic energy ratios, with amplification in the kinematic phase dominated 
by compression rather than line stretching. In contrast, at $\Pm=10$, vortical stretching 
emerges as the dominant mechanism, yielding faster growth, higher saturation levels, and 
stronger suppression of density--magnetic field correlations by magnetic pressure. This 
transition is reflected in the correlation coefficient between density and magnetic field 
strength, which is strongly positive at $\Pm=1$ but decreases significantly at higher $\Pm$. 
Across all runs, the ratio of velocity-to-magnetic integral scales is $\sim 3.4$, in the 
saturated phase, independent of $\Pm$, while the ratio of viscous to resistive dissipation 
scales increases with the increase in $\Pm$. Synthetic Faraday rotation measures reveal
coherence lengths of $\sim$one-fourth to one-third of the forcing scale across the range of $\Pm$ explored.
Using these coherence scales, we discuss the potential contribution of fluctuation dynamos 
to Faraday rotation expected from turbulent, gas-rich young disk galaxies.

\end{abstract}

\keywords{Dynamo --- Magnetohydrodynamics (MHD) --- ISM: magnetic fields --- ISM: turbulence --- Methods:Numerical}

\section{Introduction}
\label{Sec:intro} 

Magnetic fields are a fundamental component of nearly all astrophysical systems 
from stars and galaxies to the intracluster medium. Their ubiquity and persistence 
are widely attributed to dynamo processes, which amplify and sustain magnetic 
fields over cosmic time scales. Among these, the fluctuation dynamo plays a key 
role in turbulent environments, where it harnesses energy from 3D 
random/turbulent motions to efficiently amplify weak seed fields to dynamically 
significant strengths on timescales shorter than the lifetimes of such systems 
\citep{K68, R19, SS21, Scheko22}. 
This process is particularly relevant in astrophysical settings dominated 
by turbulent, conducting plasma such as star-forming regions in galaxies, and 
galaxy clusters \citep[e.g.,][]{HBD04,Scheko+04,Cho+09,Fed+11,BS13,Fed16,
XL16,XL21,Seta+20,SS24,Kriel+25}. Understanding fluctuation dynamos is 
essential for unraveling the origin and evolution of cosmic magnetism from the 
magnetic fields in the first stars \citep{Sur+10,Sur+12,FSSBK11}, and in the 
present-day interstellar and intracluster media \citep[e.g.,][]{BS13,Pak+17,
Don+18,Marin+18,SS21}, to the observational evidence for magnetization in high-redshift 
galaxies\citep[e.g.,][]{Bernet+08,Farnes+14,Geach+23,Chen+24}.

Fluctuation dynamos amplify magnetic fields through the competition between 
inductive growth and resistive dissipation. Sustained amplification requires that 
induction dominates over diffusion, which occurs only when the magnetic Reynolds 
number $\Rm$ exceeds a critical threshold, $\Rm > \Rmc$.
From the perspective of numerical simulations, this implies that both inductive and 
dissipative contributions at individual grid points tend to be significantly larger than 
their mean values. This is clearly reflected in the broad probability distribution functions 
(PDFs) presented in \citet{SS24}, where the standard deviation notably exceeds the 
mean of the PDFs. Indeed, it is the volume-integrated balance of these opposing 
effects that ultimately determines whether the magnetic field experiences net growth 
or decay.

In this study, we explore fluctuation dynamos operating in the regime of supersonic 
turbulence -- conditions highly representative of the interstellar medium (ISM) in 
galaxies. Supersonic flows introduce strong compressibility and shock-driven structures 
that can fundamentally influence the dynamics of turbulent magnetic field amplification 
\citep{KNPW07,Fed+11,Fed16,SF21a,Gent+23,Beattie+24,SS24}. 
These effects are especially pertinent in the context of early galaxy evolution and 
high-redshift environments, where the ISM is likely to be more dynamic, denser, and 
more turbulent than in present-day spiral galaxies \citep[e.g.,][]{Green+10,Bour+11,
Kraljic+24,Rizzo+24}. Fluctuation dynamos are expected to play a central 
role in such settings, acting as efficient mechanisms for the rapid amplification of 
magnetic fields from weak initial seed values. These small-scale fields may not only 
catalyse the growth of large-scale, coherent galactic magnetic fields observed today 
but could also actively facilitate their development by expelling small-scale 
magnetic helicity through helicity fluxes \citep[e.g.,][among recent works]{GS23,BV25}. 
Thus, understanding fluctuation dynamos in the supersonic regime forms a key 
component in understanding galactic magnetic fields.

We specifically explore the fluctuation dynamo in supersonic turbulence with 
rms Mach number $\mrms \approx 11$, focusing on magnetic Prandtl numbers 
$\Pm = \Rm/\Rey = \nu/\eta \gtrsim 1$, with cases explored up to $\Pm = 10$. 
Here, $\Rm$ and $\Rey$ denote the magnetic and fluid Reynolds numbers, while 
$\nu$ and $\eta$ are the kinematic viscosity and magnetic diffusivity, respectively.
Probing the $\Pm > 1$ regime is particularly relevant for the ISM, as such systems are 
expected to have $\Pm \gg 1$, with estimates for the warm diffuse phase reaching
$\Pm \sim 10^{10} - 10^{14}$ \citep{BS05,Schober+12}. However, we restrict our 
exploration to $\Pm \leq 10$ to maintain computational feasibility. Within this 
parameter space, we seek to address some fundamental questions from the 
perspective of the $\Pm = 1$ and $\Pm > 1$ regimes. By contrasting these two 
regimes, we first explore how $\Pm$ influences the growth rate and saturation 
levels of magnetic energy. Next, how does the correlation between density and 
magnetic field strength evolve across different $\Pm$ values, and what roles do 
the weak and strong field regions play in shaping this correlation? We further 
investigate the balance between local stretching and compression during the 
saturated phase of the dynamo and assess the resulting magnetic field coherence 
using synthetic Faraday rotation measures (RMs). Together, these analyses provide 
new insights into the nature and efficiency of fluctuation dynamos. 

The paper is structured as follows. Section~\ref{s:nsims} details the numerical 
setup, including initial and boundary conditions. Section~\ref{s:time_series} presents 
visualizations of density and magnetic field strengths, along with the time evolution 
of $\mrms$ and the magnetic-to-kinetic energy ratios. In Section~\ref{s:corr}, we 
examine the correlation between density and magnetic field strengths, highlighting 
differences across weak and strong field regions. Section~\ref{s:scales} discusses 
the power spectra and characteristic length scales, such as the integral and dissipation 
scales of the velocity and magnetic fields. Using the magnetic energy evolution as 
a starting point, Section~\ref{s:energy} investigates the roles of local stretching and 
compression and their dependence on $\Pm$. Section~\ref{s:faraday} is devoted to 
an analysis of the Faraday RM obtained from the dynamo-generated fields, in the different physical regimes quantified by the varying $\Pm$. 
Finally, in Section~\ref{s:conc} we summarize the key findings and discuss their 
implications.

\section{Numerical Simulations}
\label{s:nsims}

We perform nonideal magnetohydrodynamic (MHD) simulations of fluctuation dynamos 
in supersonic flows in three dimensions using a newly developed driven turbulence 
module in PLUTO \citep{M+07}\footnote{\url{https://plutocode.ph.unito.it/}}, which is a 
widely used, finite-volume, astrophysical MHD code. 
The simulations were performed in dimensionless coordinates with a cubic box of unit 
length ($L = 1$) at $512^{3}$ resolution. The dimensionless density ($\rho$) and sound 
speed ($c_{\rm s}$) are initialized with $\rho = 1, c_{\rm s} = 1$ and zero initial velocities. 
Adopting an isothermal equation of state, we solve the following set of 3D MHD equations 
in dimensionless form:
\begin{align}
&\frac{\partial \rho}{\partial t} + \nabla \cdot (\rho \UU) = 0,  \label{eq:ce} \\
&\frac{\partial (\rho \UU)}{\partial t} + \nabla \cdot \left(\rho\UU \otimes \UU - \BB \otimes \BB\right) 
+ \nabla P^{*} = \nabla \cdot (2 \nu \rho \SSS) + \rho \FF, \label{eq:ns} \\
&\frac{\partial \BB}{\partial t} = \nabla \times (\UU\times \BB) + \eta \nabla^2 \BB.  \label{eq:ie} 
\end{align}

Here $\rho, \UU, P^{*} = p + |\BB|^{2}/2$ and $\BB$ represent the fluid density, velocity, total 
pressure (thermal + magnetic) and magnetic field, respectively, while $\otimes$ denotes the 
tensor product between vector fields. Furthermore, $S_{ij} = (1/2)[U_{i,j} + U_{j,i}  -(2/3)\delta_{ij}\partial_{k}U_{k}]$ 
is the traceless rate of the strain tensor, and $\FF$ is the turbulent acceleration field modeled 
using the Ornstein-Uhlenbeck process with a finite time correlation \citep{EP88,Gill96,
Fry+00, Benzi+08, Fed+08}. The viscosity $\nu$ and magnetic resistivity $\eta$ are both 
treated as constants throughout the simulations. 

To regulate solenoidal and compressive contributions in our driven turbulence module, we 
decompose the acceleration field into solenoidal and compressive components using a 
projection operator in Fourier space. In index notation, the operator is given by,
\begin{align}
\mathcal{P}^{\zeta}_{ij}(\kk) = \zeta\mathcal{P}^{\perp}_{ij} + (1-\zeta)\mathcal{P}^{\parallel}_{ij},
\end{align}
where $\mathcal{P}^{\perp}_{ij}$ and $\mathcal{P}^{\parallel}_{ij}$ are the solenoidal and 
compressive projection operators, respectively, and $\zeta \in [0,1]$ is an adjustable parameter
that controls the solenoidal contribution. To maximize dynamo efficiency in supersonic 
turbulence, we use purely solenoidal driving (choosing $\zeta = 1$)\footnote{$\zeta = 0$ 
implies purely compressive driving, and any value in between implies mixed driving.} in all 
simulations ensuring $\kk\cdot \FF_{k} = 0$, ($k$ is the wavenumber and $\FF_{k}$ is the 
forcing vector in $k$-space), exciting only large-scale modes in the range $1 \leq |\kk|L/2\uppi \leq 3$, 
with average forcing wavenumber $k_{\rm f}L/2\uppi = 2$, corresponding to turbulent driving 
scale $\lf = 2\uppi/\kf = L/2$. We further set the correlation time to be the eddy-turnover time 
at this scale, $\ted = \lf/\urms$, where $\urms$ is the steady-state rms turbulent velocity. The 
amplitude of the driving is adjusted to yield supersonic turbulence with rms Mach number 
$\mrms = \urms/c_{\rm s} \approx 11$ when the turbulence is fully developed. 

We initialize the setup with a magnetic field $\BB = B_{0}[(\sin(15\pi z), 0, 0)]$, where $B_{0}$ 
is chosen so that the initial plasma beta $\beta_{\rm in} = p_{\rm th}/p_{\rm mag} \approx 10^{6}$ 
in all the runs; $p_{\rm th}$ and $p_{\rm mag}$ are the thermal and magnetic pressures, 
respectively. Equations~(\ref{eq:ce}) -- (\ref{eq:ie}) are then evolved with an explicit time 
stepping scheme together with the unsplit staggered mesh MHD Harten, Lax, Van Leer (HLL) solver to compute the 
fluxes and a constrained transport scheme at the cell interfaces for preserving the 
divergence-free nature of the magnetic fields ($\nabla\cdot\BB =0$) on the staggered grid. We 
note that even though the ISM exhibits a multiphase structure, the use 
of an isothermal equation of state enables a cleaner examination of the complex interplay of 
density fluctuations, turbulence, and magnetic fields. To achieve $\Pm = \Rm/\Rey = \nu/\eta > 1$, 
we progressively reduce $\Rey$ by increasing $\nu$, resulting in 
$\Rey = \urms\lf/\nu \approx 6600,1320$ and $660$ for $\Pm = 1, 5$ and $10$, respectively. 
The key simulation parameters are listed in Table~\ref{tab:sumsim}. For clarity, we will hereafter 
refer to the different runs as Pm1, Pm5, and Pm10, corresponding to $\Pm = 1, 5$ and $10$, 
respectively.

\begin{table}
\centering 
\setlength{\tabcolsep}{4pt} 

\caption{Key parameters of simulations used in This Study}
\begin{tabular}{ccccc} \hline
Run & $\Pm$ & $\Rey = u\,\ell_{\rm f}/\nu$ & $\langle E_{\rm m}/E_{\rm k}\rangle_{\rm sat}$
& $\langle r_{p}\rangle_{\rm sat}$ \\ \hline
Pm1 & 1 & 6600 & $9.0\times 10^{-3} \pm 4\times 10^{-4}$ & $0.53 \pm 0.02$\\  
Pm5 & 5 & 1320 & $3.5\times 10^{-2} \pm 2\times 10^{-3}$ & $0.45 \pm 0.01$ \\  
Pm10 & 10 & 660 & $8.1\times 10^{-2} \pm 2\times 10^{-3}$ & $0.34 \pm 0.01$ \\ \hline
\end{tabular}
{\raggedright \textbf{Note.} The resolution in each run is $512^{3}$. $k_{\rm f}L/2\pi = 2 $ is the average forcing 
wave number and $\mrms \approx 11$ is the average value of the rms Mach 
number in the steady state. $\ell_{\rm f} = 2\pi/k_{\rm f}$ is the 
forcing scale and $\Pm$ and $\Rey$ are the magnetic Prandtl number 
and the fluid Reynolds numbers, respectively. The ratio of the time-averaged 
magnetic to kinetic energies $\langle E_{\rm m}/E_{\rm k}\rangle$, and the 
correlation coefficient $\langle r_{p}\rangle$ are computed in the saturated 
state of the dynamo. The $\pm$ values indicate the $1\sigma$ standard
deviation around the mean. \par}
\label{tab:sumsim}
\end{table}

\section{2D Slices and Time Evolution of the rms Mach Number and Ratio of Energies}
\label{s:time_series}

\begin{figure*}
\includegraphics[width=\textwidth]{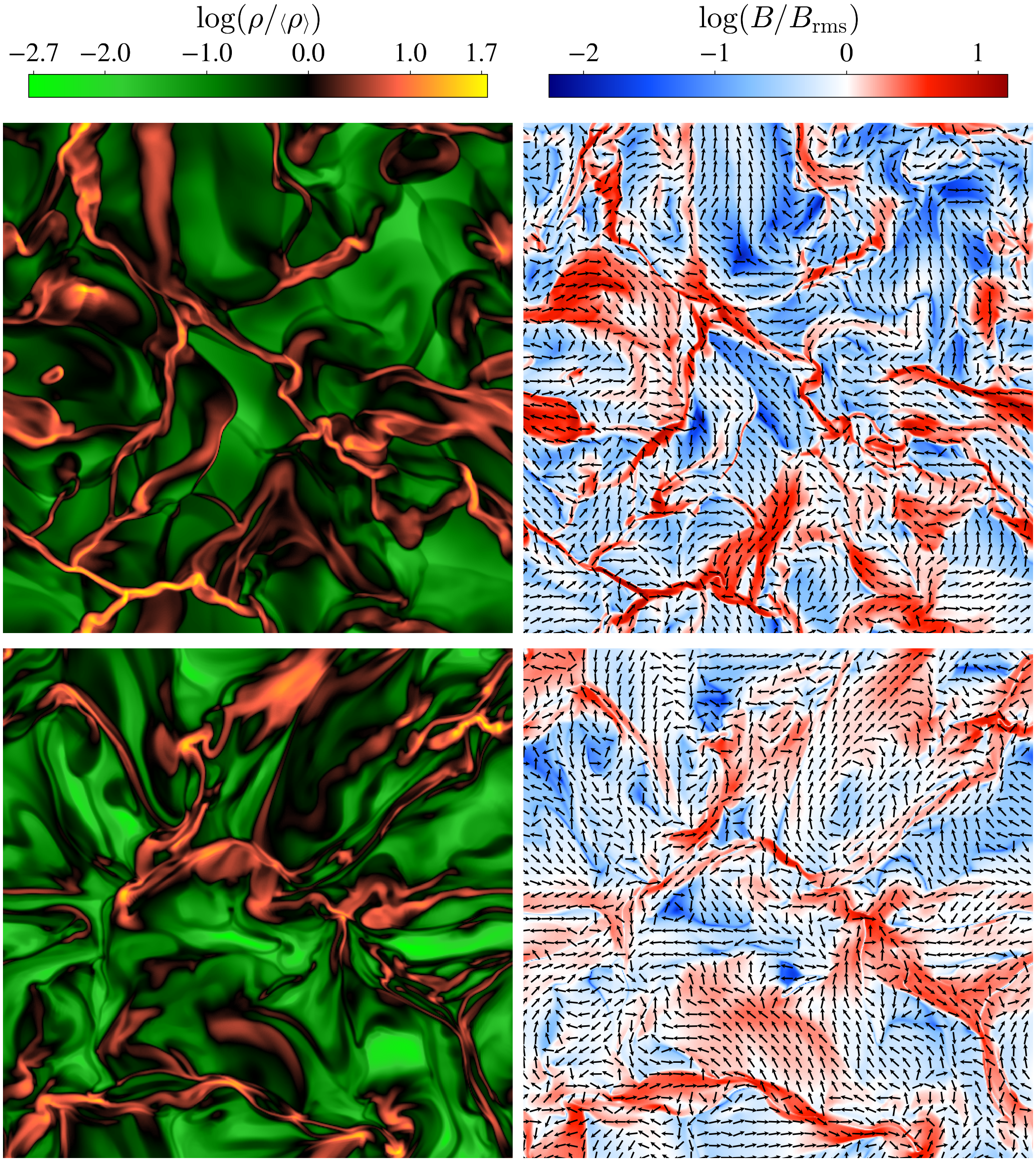}
\caption{2D slices in the $x-z$ plane at $y=0.5$ from Pm1 (top row) 
and Pm10 (bottom row) in the saturated phase. The left panels show logarithmic density 
contrasts, $\log(\rho/\langle\rho\rangle)$ with the color scale ranging from low-density 
voids (green) to highly dense structures (red and yellow). The right panels display the 
logarithmic values of the normalized field strength, $\log(B/\brms)$, with a blue-to-red 
color scale depicting regions of low-to-high field strengths. The arrows of equal length 
represent the directions of the in-plane magnetic field vectors.}
\label{fig:2d_slices}
\end{figure*}

In Figure~\ref{fig:2d_slices}, we show the 2D slices of the logarithmic values of the normalized 
density and magnetic field strengths from Pm1 (top row) and Pm10 (bottom row) in the 
saturated state of the dynamo. They depict the complex structure of a magnetized, 
compressible turbulent medium. The left-hand panels visualize $\log(\rho/\langle\rho\rangle$), 
using a color map ranging from green (low density, $\log(\rho/\langle\rho\rangle \approx -2.7$) 
to yellow-white (high density, $\log(\rho/\langle\rho\rangle \approx 1.7$). Here 
$\langle\rho\rangle \approx 1$ is the mean density in the simulation volume. It is abundantly 
clear that the supersonic nature of the resulting flows results in sharp density contrasts with 
high-density structures surrounded by underdense voids. Such structures are hallmark 
features of compressible turbulence, particularly in supersonic regimes, where shock fronts 
compress the gas into thin layers. 

The right-hand panels show the magnetic field intensity, $\log(B/\brms)$, overlaid with vector 
field arrows representing the in-plane field direction. Blue regions denote weak 
magnetic fields (sub-rms values), while red regions indicate strong magnetic concentrations 
exceeding $\brms$. These slices reveal that, irrespective of the value of $\Pm$, regions with 
high densities generally correspond to regions with strong magnetic fields, and vice versa, 
with sub-rms field strengths in the voids. However, a lack of perfect correlation between 
strong field and high-density regions is also visible in some areas. The arrows of equal 
length representing the direction of the in-plane fields are seen to be arranged in folds at a 
number of locations. Some highly dense blobs show ordered magnetic field lines plausibly 
resulting from compression. This is expected, as in purely solenoidal supersonic turbulence, 
both compression and random stretching contribute to amplifying the magnetic field through 
dynamo action \citep{Fed+11, Fed16, SBS18, SF21a,SS24}. 

\begin{figure}[h!]
\begin{centering}
\includegraphics[width=\columnwidth]{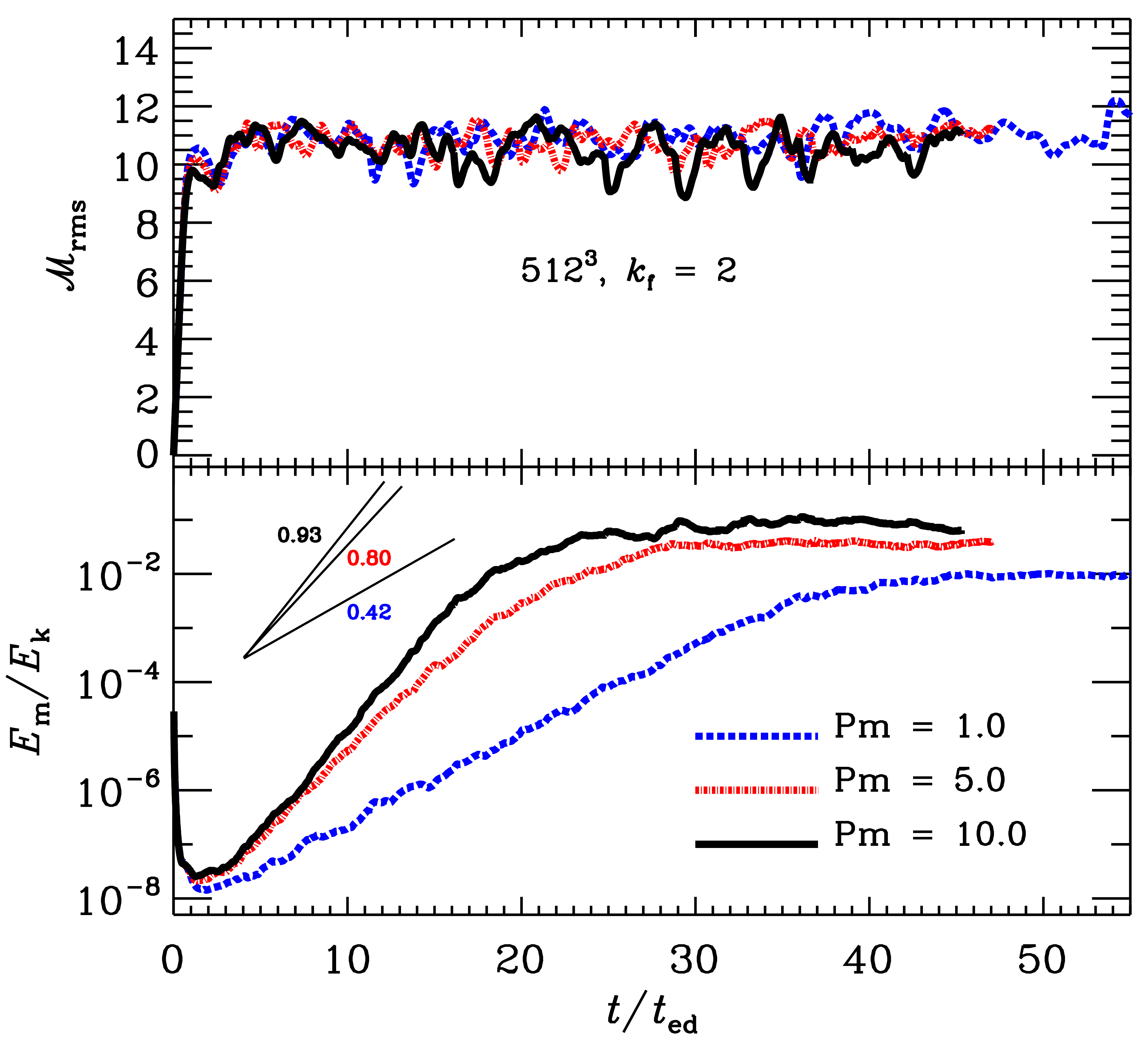}
\end{centering}
\caption{Evolution of $\mrms$ (top) and $E_{\rm m}/E_{\rm k}$ (bottom) with $t/\ted$ 
for Pm1 (blue dashed), Pm5 (red dotted) and Pm10 (black solid). While $\mrms \approx 11$ 
in the steady state, the growth rates in the kinematic phase and the saturation level of 
$E_{\rm m}/E_{\rm k}$ are strongly dependent on the $\Pm$, with higher values producing 
faster exponential growth and higher saturation levels. The annotated slopes indicate the 
growth rates in the kinematic phases.}
\label{fig:tseries_mrms_emag}
\end{figure}

The top panel in Figure~\ref{fig:tseries_mrms_emag} shows the evolution of the rms Mach 
number ($\mrms$), while the bottom panel shows the evolution of the ratio of magnetic to 
kinetic energies ($E_{\rm m}/E_{\rm k}$), both as functions of the normalized time ($t/\ted$). 
The behavior of $\mrms$ is shown for three different $\Pm$ values: $\Pm = 1$ (blue 
dashed line), $\Pm = 5$ (red dotted line), and $\Pm = 10$ (solid black line). As seen from 
the plot, there is an initial transient phase lasting for $\approx 2 t/\ted$ beyond which $\mrms$ 
attains a steady state fluctuating around $\mrms\approx 11$, obtained by averaging from 
$t/\ted = 5$, till the end of the simulation. The growth of magnetic fields due to dynamo 
action proceeds in three stages: kinematic, intermediate, and eventual saturation. We find 
that the behavior of the fluctuation dynamo in such highly compressible flows to be sensitive 
to the relation between the ranges of scales of velocity and magnetic fields, quantified 
in terms of $\Pm = \nu/\eta$. Notably, the growth rate of the dynamo varies significantly 
with $\Pm$. In the kinematic phase, these are annotated by slopes in the top left corner 
of the bottom panel. In Pm1, the growth rate of $E_{\rm m}/E_{\rm k}$ is significantly 
lower ($\gamma \approx 0.42 \pm 0.001$) than Pm5 ($\gamma \approx 0.8 \pm 0.003$) 
and Pm10 ($\gamma \approx 0.93 \pm 0.002$). 
This implies that in contrast to $\Pm = 1$, $\Pm = 10$ results 
in more efficient dynamo action, as the magnetic fields can be amplified on a range of 
scales ($\ell_{\nu} > \ell > \ell_{\eta}$), even by the eddies at the viscous scale. 
The intermediate phase of growth is particularly prominent in Pm10 extending 
from $t/\ted = 19$ to $25$, while for Pm1 and Pm5 it lasts from $t/\ted = 36$ to $41$ and 
$21$ to $28$, respectively. In the same vein, we find that the saturation level of $E_{\rm m}$ 
depends strongly on the $\Pm$, with the highest level of saturation obtained for Pm10 
($\approx 8\times10^{-2}$), which decreases with decreasing $\Pm$ (see Table~\ref{tab:sumsim}). 
Further, the $1\sigma$ uncertainties in the saturated $E_{\rm m}/E_{\rm k}$ values 
reported in Table~\ref{tab:sumsim} correspond to only $\sim 4\%-6\%$ of the mean.

\section{Correlation between density and magnetic field strength}
\label{s:corr}

In a recent work, \citet{SS24} explored how the Pearson correlation coefficient, $r_{p}(\rho,B)$
evolves with varying levels of flow compressibility. The coefficient is defined, as
\begin{align} 
r_{p}(\rho, B) = \frac{{\rm Cov}(\rho, B)}{\sigma_{\rho}\sigma_{B}} &  \\ & \hspace{-2cm}\nonumber
= \frac{\sum_{i,j,k}(\rho_{i,j,k} - \mrho)(B_{i,j,k} - \mbmag)}
{\sqrt{\sum_{i,j,k}(\rho_{i,j,k} - \mrho)^{2}}\,\sqrt{\sum_{i,j,k}(B_{i,j,k} - \mbmag)^{2}}},
\label{rp}
\end{align}
where $B = \sqrt{B_x^{2} + B_{y}^{2} + B_{z}^2}$ is the magnitude of the field and $\rho_{i,j,k}$ 
and $B_{i,j,k}$ are the density and the magnetic field strength at a point $(i,j,k)$ in the 
simulation volume. $\mrho$ and $\mbmag$ are the mean values of density and $B$, respectively. 
One of the key findings from \citet{SS24} was that, in supersonic flows with $\mrms \sim 3$, 
$\rho$ and $B$ remain positively correlated even in the nonlinear saturated regime, with 
$\langle r_{p}\rangle \approx 0.43$. In this study, we further examine the evolution of 
$r_{p}(\rho, B)$ across different $\Pm$ values in flows with $\mrms \sim 11$. 

\begin{figure}[t!]
\begin{centering}
\includegraphics[width=\columnwidth]{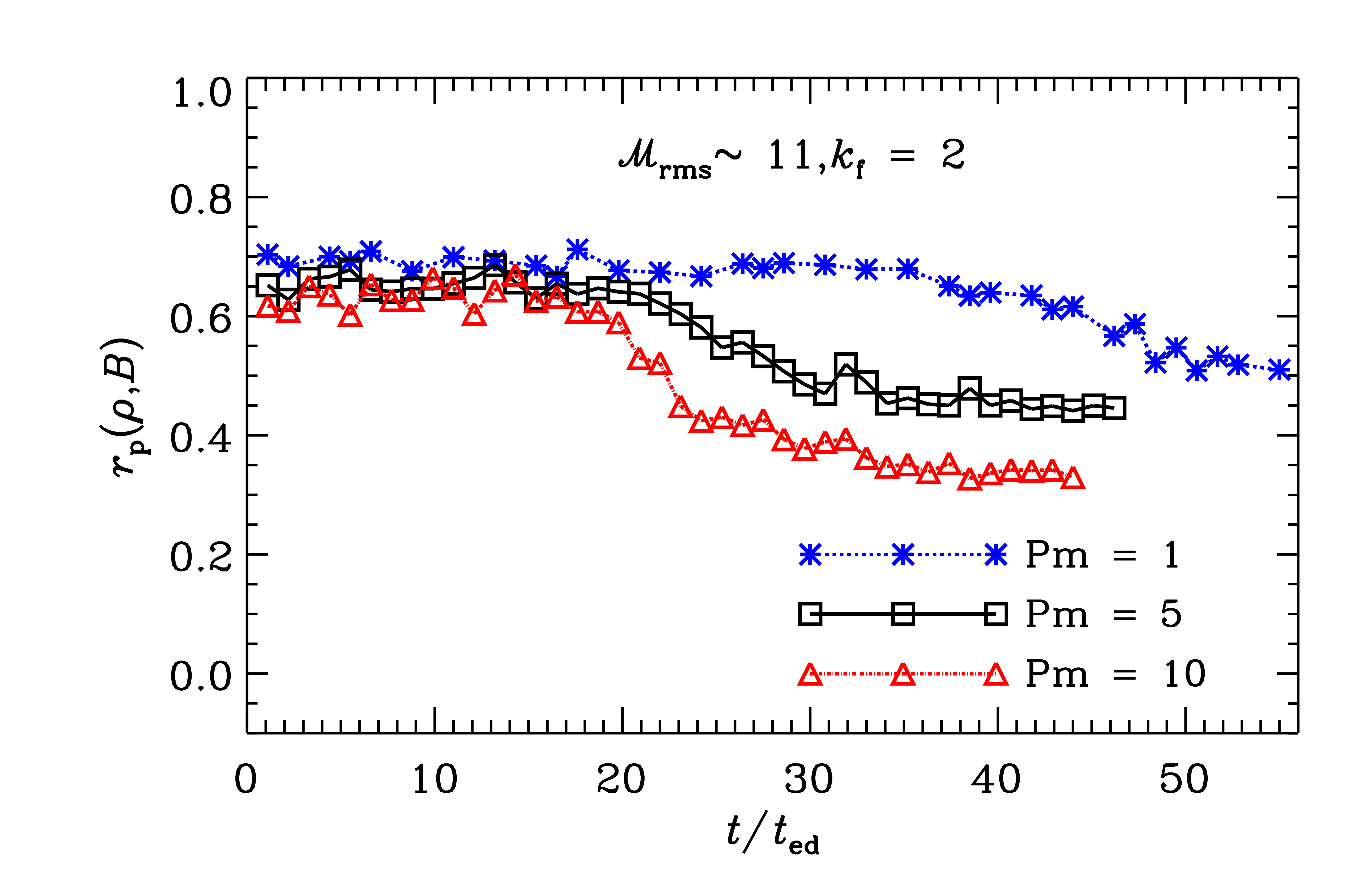} 
\end{centering}
\caption{Evolution of the Pearson correlation coefficient $r_{p}(\rho, B)$ with $t/\ted$ 
for Pm1 (blue asterisks), Pm5 (black squares), and Pm10 (red triangles). Starting from 
a strong positive correlation, $r_{p}(\rho,B)$ drops as the dynamo evolves from the 
kinematic to saturated phase. The decrease in $r_{p}(\rho, B)$ is strongly dependent 
on $\Pm$, with the $\Pm = 10$ run decreasing by $\approx 52\%$ from its value in the 
kinematic phase.}
\label{fig:corr_m11_diffpm}
\end{figure}

During the kinematic phase, it is evident from Figure~\ref{fig:corr_m11_diffpm} that all 
three simulations (Pm1, Pm5, and Pm10) exhibit a strong positive correlation with 
time-averaged $\langle r_{p}\rangle \approx 0.65$ indicating that initially, higher-density 
regions are correlated with stronger magnetic fields. However, as the dynamo transitions 
to the nonlinear saturated phase, $r_{p}$ starts to decrease. This decline stems from the 
fact that magnetic field amplification is no longer driven solely by compression; vortical 
motions arising from solenoidal forcing also contribute significantly to field growth via 
random stretching. We find the weakening of the correlation to be more pronounced at 
high $\Pm$. The steady-state values of $r_{p}$ averaged over multiple independent 
realizations of the saturated state across different runs together with their $1\sigma$ 
variations are listed in Table~\ref{tab:sumsim}. For Pm1, $\langle r_{p}\rangle$ is computed 
using eight realizations between $t/\ted = 46-55$ and $12$ realizations each for both 
Pm5 and Pm10 covering $t/\ted = 32-46$ and $t/\ted = 32-45$, respectively. We find that
while $\langle r_{p}\rangle$ attains values of $\approx 0.53$ and $\approx 0.46$ for Pm1 
and Pm5, it declines steeply, settling to $\approx 0.34$ for Pm10. Notably, \citet{SF21b} 
obtained $\langle r_{p}\rangle \approx 0.56 \pm 0.02$ in their $\mrms \approx 10, \Pm = 1$ 
simulation, which is very close to our Pm1 value.

The observed decrease in the degree of positive correlation in Pm10 likely highlights 
the role of magnetic pressure forces, as discussed in the Appendix~\ref{sec:align}. This interpretation 
is further supported by Fig.~\ref{fig:align} which shows the PDFs of the cosine of the angle 
between the unit vector of the gradient of the density (${\bf n}_{\nabla\rho}$) and the gradient 
of magnetic pressure (${\bf n}_{\nabla B^{2}}$). The figure clearly demonstrates that the 
antiparallel alignment between $\nabla\rho$ and $\nabla B^{2}$ strengthens in the saturated 
phase for $\Pm > 1$, while the parallel alignment weakens. This suggests that magnetic 
pressure forces oppose further compression of the field lines in the nonlinear saturated phase. 
As a result, density variations become more anticorrelated with variations in magnetic pressure 
at $\Pm > 1$, leading to a steep decline in $r_{p}$. 

\subsection{Evolution of $r_{p}(\rho, B)$ in regions of strong/weak fields} 
\label{ssec:diff_field}

\begin{figure}[t!]
\begin{centering}
\includegraphics[width=\columnwidth]{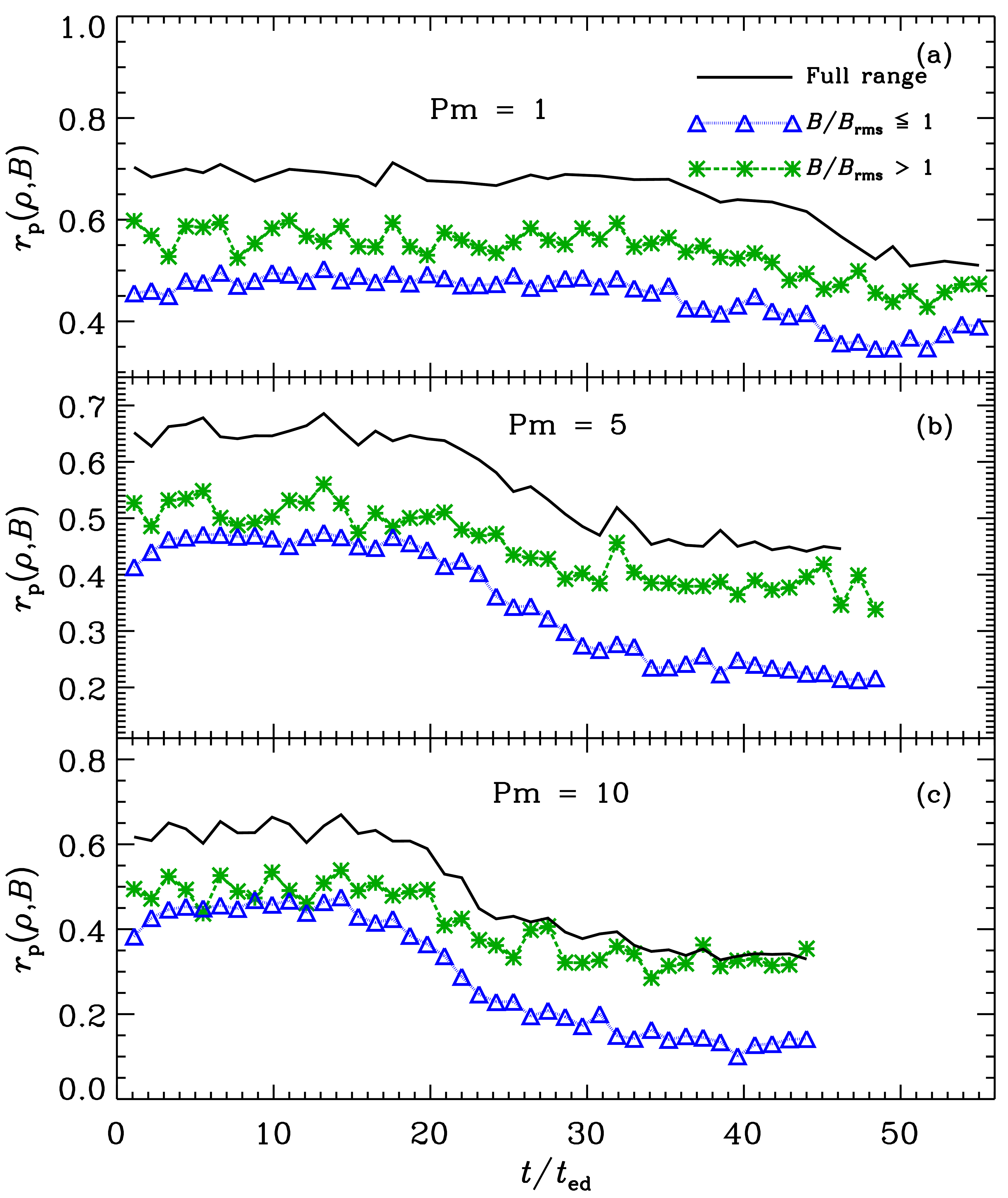} 
\end{centering}
\caption{Evolution of $r_{p}(\rho, B)$ for Pm1 (top), Pm5 (middle), and Pm10 (bottom) in 
different ranges of $B/\brms$. The solid black line shows the evolution over the full range 
of magnetic field strengths similar to Fig.~\ref{fig:corr_m11_diffpm}. The blue open triangles
represent $r_{p}(\rho,B)$ in regions where $B/\brms \leq 1$, while green asterisks denote 
correlations in regions with $B/\brms > 1$. 
}
\label{fig:corr_m11_bcuts}
\end{figure}

To gain insight into the observed decrease of $r_{p}(\rho, B)$ with increasing 
$\Pm$, we examine how this correlation evolves within regions of differing 
magnetic field strength, specifically, areas where $B/\brms \leq 1$ (weak fields) 
and $B/\brms > 1$ (strong fields). This analysis is presented in 
Figure~\ref{fig:corr_m11_bcuts}, with panels (a), (b), and (c) corresponding to 
runs Pm1, Pm5, and Pm10, respectively. It is important to highlight that fluctuation 
dynamos naturally generate intermittent magnetic fields with a broad distribution 
of strengths, including regions with $B/\brms>1$, even in incompressible turbulence, 
where amplification arises purely from random stretching. The supersonic regime 
studied here introduces an additional amplification due to compression that further 
broadens the high-$B$ tail \citep{SS24}. Thus, the strong field structures visible in 
Figure~\ref{fig:2d_slices} arise from the combined effects of compression and 
random stretching, with compression enhancing but not uniquely producing the 
regions with $B/\brms>1$.

In each panel, the solid black line depicts the evolution of $r_{p}(\rho,B)$ where no range 
of $B/\brms$ is considered. These lines correspond to those in Fig.~\ref{fig:corr_m11_diffpm} 
and are presented here for an effective comparison. 
In the top panel, similar to the declining trend observed for the black solid curve, $r_{p}(\rho, B)$
also shows a declining trend in the strong field regime (green asterisks), although the correlation 
is consistently lower, fluctuating between $0.5$ and $0.6$ before settling to $\approx 0.45$ in 
the saturated state of the dynamo. For the weak field regime (blue triangles), the correlation is 
lower overall, starting around $0.5$ and dropping steadily after $t/\ted\approx 30$, reaching 
a steady-state value of $\approx 0.36$ by the end of the run. This suggests that at $\Pm = 1$, 
$\rho$ and $B$ are moderately correlated, particularly in stronger field regions, but the 
correlation weakens over time, especially for weaker magnetic fields. In fact, even though 
$r_{p}$ decreases by a similar factor in both strong and weak-field regions, the overall higher 
value of $r_{p}$ in the strong field regions implies that the compression of the fields due to 
density enhancements still dominates the $\rho-B$ correlation. 

For $\Pm = 5$, panel (b) shows that strong field correlation (green asterisks) starts near 
$0.5$ and decreases more gradually to a steady-state value of $\approx 0.39$. In comparison, 
$r_{p}$ for weak field regions (blue triangles) begins just below $0.5$ but drops more 
steeply to values $\approx 0.24$. The rate of decline is more pronounced than in the 
$\Pm = 1$ case, indicating stronger dissociation between $\rho$ and $B$ as the dynamo 
evolves to the saturated state. Finally, for $\Pm = 10$,  panel (c) shows that the trends 
continue with further reduction in $r_{p}(\rho,B)$. In this case, the strong field correlation 
(green asterisks) remains relatively higher than the weak field (blue triangles), with values 
fluctuating between $0.45$ and $0.55$ before settling to a steady-state value of $\approx 0.32$. 
On the other hand, $r_{p}(\rho,B)$ in regions with $B/\brms \leq 1$ shows the most notable 
decline, starting at approximately $0.45$ and reducing by a factor $\approx 3$ to settle at 
a steady-state value of $\approx 0.14$. This reduction in $r_{p}$ in the weak field regions 
is stronger compared to a decline by factor of $\approx 1.5$ in the strong field regions.

In summary, a clear trend emerges across all three runs: as the $\Pm$ increases, the 
overall correlation between $\rho$ and $B$ decreases. The trend seen for $\Pm = 1$ 
is similar to the findings in \citet{SS24} for $\mrms = 3$. This decrease could be due to 
the fact that as magnetic fields grow in importance, forces due to magnetic pressure 
start to resist the further compression of field lines \citep{SS24}. However, the $\Pm$ 
dependence of this decrease adds a new perspective to our results. In particular, 
the fact that the decrease in $r_{p}$ is relatively minor in the strong field regions suggests 
that fields in those regions are constantly being amplified due to compressions in 
supersonic flows, which compensate for the magnetic pressure forces. On the other hand, 
the general sea of volume-filling, $B/\brms \leq 1$ fields mainly arises due to random 
stretching. In the absence of strong compression in these regions, magnetic pressure 
forces manage to dominate over compression and reduce $r_{p}$. This effect is stronger 
at $\Pm = 10$ compared to $\Pm = 1$, due to efficient dynamo action in the former.

\section{Power spectra and characteristic scales}
\label{s:scales} 

\begin{figure}[t!]
\begin{centering}
\includegraphics[width=\columnwidth]{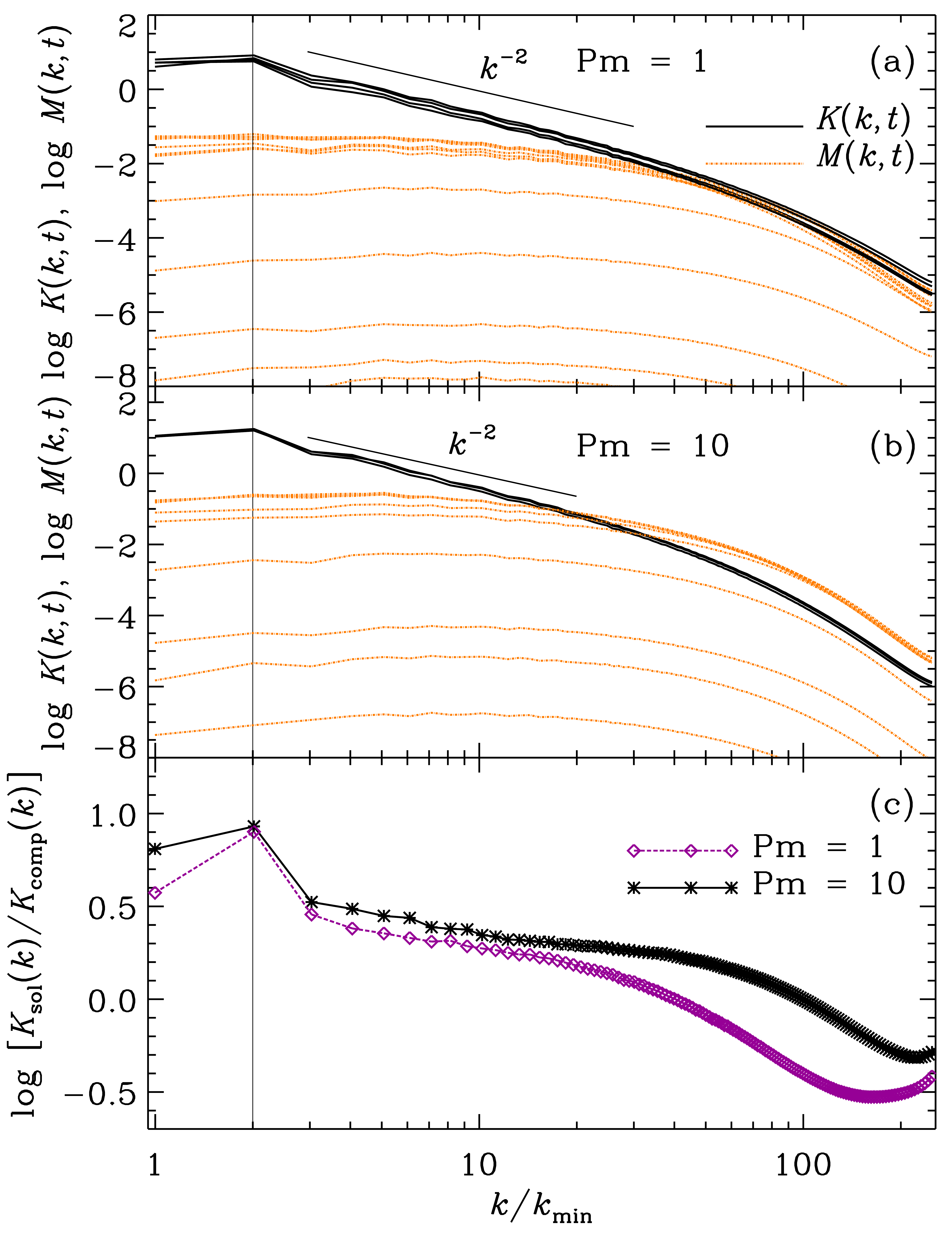} 
\end{centering}
\caption{Kinetic $K(k,t)$ and magnetic $M(k,t)$ energy spectra for Pm1 and Pm10 at fixed $\Rm$. 
Panels (a) and (b) show the time evolution of $K(k,t)$ (solid 
black lines) and $M(k,t)$ (dotted orange lines) as functions of wavenumber $k/k_{\rm min}$ 
for $\Pm = 1$ and  $\Pm =10$, respectively. Panel (c) shows the time-averaged ratio of 
solenoidal to compressive components of the kinetic energy spectrum, $K_{\rm sol}(k)/K_{\rm comp}(k)$, 
for both runs in the saturated phase of the dynamo. Higher $\Pm$ leads to a relatively larger 
solenoidal energy fraction at intermediate and small scales and enhances magnetic energy 
at small scales compared to the $\Pm = 1$ case. The thin vertical at $k/k_{\rm min} = 2$ is 
the turbulence driving scale, where $k_{\rm min} = 2\pi\,L^{-1}$ is the smallest wavenumber 
in the box. The $K(k,t)$ shown here are density-weighted velocity spectra ($\sqrt{\mrho}\UU$). 
Since $\mrho \approx 1$ in our simulation volume, the weighting does not affect the spectral 
slopes.
}
\label{fig:spectra}
\end{figure}

\subsection{Power Spectra}
\label{ssec:power_spec}

Panels (a) and (b) in Fig.~\ref{fig:spectra} show the time evolution of the 1D 
shell integrated kinetic $K(k,t)$ and magnetic $M(k,t)$ spectra for runs Pm1 and Pm10, 
respectively. Here $K(k,t)$ is computed by weighting the velocity components by 
$\sqrt{\langle\rho\rangle}$. In both cases, we find that the slope of $K(k,t) \approx k^{-2.1}$, 
slightly steeper than the $k^{-2}$, expected in hydrodynamic supersonic turbulence. 
The evolution of $M(k,t)$ is shown at different times from the kinematic to the saturated 
phase. In agreement with previous works \citep{Fed16,SF21a,SS24}, $M(k,t)$ evolves in 
a self-similar fashion that points to the nonlocal feature of the fluctuation dynamo in 
$k$-space \citep{SS21}. In Pm1, we find that by $t/\ted = 55$, the magnetic energy is still 
less than the kinetic energy on all but very small scales beyond $k/k_{\rm min} \sim 40$. 

The bottom panel in Fig.~\ref{fig:spectra} shows the time-averaged spectra of the ratio of 
solenoidal to compressive kinetic energy, $K_{\rm sol}(k)/K_{\rm comp}(k)$, for Pm1
(dashed, diamonds) and Pm10 (solid, asterisks) in the saturated phase. This is 
obtained by first decomposing the velocity field into a divergence-free ($\UU_{\rm sol}$) 
and a curl-free ($\UU_{\rm comp}$) components using a Hodge-Helmholtz decomposition, 
followed by evaluating the spectra of $K_{\rm sol}(k)$ and $K_{\rm comp}(k)$. Recall that 
in our simulations, the $\Rm$ is held fixed, so increasing $\Pm$ implies increasing $\nu$
while keeping $\eta$ constant. This choice isolates the effects of viscous dissipation 
on the velocity field, which in turn affects the solenoidal and compressive components 
differently. 

For example, at high Pm (e.g., Pm10), the viscosity is large, which selectively damps 
out small-scale velocity fluctuations -- but crucially, this damping is stronger for compressive 
motions, which involve velocity divergence and hence strong density and pressure gradients. 
On the other hand, solenoidal motions, being divergence-free, do not generate such gradients 
and are more resilient against viscous damping. As a result, in Pm10, compressive motions 
are preferentially suppressed at small-to-intermediate scales, leading to a higher 
solenoidal-to-compressive energy ratio over much of the inertial and dissipation ranges. 
Thus, the black solid curve($\Pm = 10$) lies well above the purple dotted ($\Pm = 1$) curve across 
a wide range of $k$. 

In contrast, for Pm1, viscosity and resistivity are the same. The flow experiences less 
viscous damping overall, and compressive motions can survive and even thrive in the highly 
supersonic regime (with $\mrms \approx 11$). Shocks and compressive features are naturally 
produced in such flows, and when not heavily damped by viscosity, they maintain a significant 
fraction of the kinetic energy budget. Hence, the solenoidal-to-compressive energy ratio is 
lower for Pm1, as seen in the purple curve\footnote{The rise of $K_{\rm sol}(k)/K_{\rm comp}(k)$ 
at $k/k_{\rm min} > 100$ reflects resolution-dependent numerical artifacts confined to the 
dissipation range and does not affect the inertial-range results.}.

\subsection{Characteristic scales from power spectra}
\label{ssec:char_scales}

Using the energy spectra $K(k,t)$ and $M(k,t)$, we first estimate the integral scales of the 
velocity and magnetic fields, defined as, 
\EQ
\label{int_scales}
\ell^{V}_{\rm int} = \frac{2\pi\int [K(k,t)/k]\,dk}{\int K(k,t)\,dk}, \,\,\,\,\,
\ell^{M}_{\rm int} = \frac{2\pi\int [M(k,t)/k]\,dk}{\int M(k,t)\,dk}, 
\EN

Across all runs (Pm1, Pm5, and Pm10), we find that $\ell^{V}_{\rm int}$ remains nearly 
constant as the dynamo evolves, whereas $\ell^{M}_{\rm int}$ grows by a factor of 
$\sim 2$ from the kinematic to the saturated phase. This is due to the influence of 
Lorentz forces that leads to larger coherence scale of the magnetic field. The third 
column of Table~\ref{tab:lscales} lists the time-averaged ratio of 
$\ell^{V}_{\rm int}/\ell^{M}_{\rm int}$ in the saturated phase, using nine realizations 
for Pm1 and $12$ realizations each for the Pm5 and Pm10 runs.
In all cases, $\ell^{V}_{\rm int}$ exceeds $\ell^{M}_{\rm int}$ by a factor of $\sim 3.4$, 
implying that even after saturation, most magnetic energy remains at $k > \kf$. 
The independence of this ratio on $\Pm$ likely arises from fixing $\Rm$ while 
reducing $\Rey$ to realize $\Pm > 1$.

\begin{table}
\centering 
\setlength{\tabcolsep}{10.0 pt}
\caption{Time-averaged Values and $1\sigma$ Variations of the Ratio of Different Characteristic 
Scales in the Saturated Phase.}
\begin{tabular}{cccc} \hline
Run & $\Pm$ & $\langle\ell^{V}_{\rm int}/\ell^{M}_{\rm int}\rangle_{\rm sat}$ & 
$\langle\ell_{\nu}/\ell_{\eta}\rangle_{\rm sat}$ \\ \hline
Pm1 & 1 & $3.40 \pm 0.20$ & $1.36 \pm 0.05$ \\  
Pm5 & 5 & $3.43 \pm 0.18$ & $1.70 \pm 0.03$ \\  
Pm10 & 10 & $3.40 \pm 0.15$ & $1.92 \pm 0.05$ \\ \hline
\end{tabular}
\label{tab:lscales}
\end{table}

We now turn to evaluating the viscous and resistive dissipation scales of the velocity 
and magnetic fields from the spectra $K(k,t)$ and $M(k,t)$. Commonly, the characteristic 
dissipation wavenumbers of the velocity fields are computed from the maximum of the 
dissipation spectra, $k_{\nu} = [\int\,k^{2}K(k)\,dk/\int K(k)\,dk]^{1/2}$ (and analogously 
for magnetic dissipation). However, this approach may be unreliable, as the location of 
the peak can be highly sensitive to the localized spectral features. Hence, to obtain a 
more robust measure, we define the viscous and resistive dissipation wavenumbers 
as
\begin{align}
\label{diss_scales}
k_{\nu} = \frac{\int\,k\,[k^{2}\,K(k,t)]\,dk}{\int k^{2}\,K(k,t)\,dk}, \,\,\,\,\,
k_{\eta} = \frac{\int\,k\,[k^{2}\,M(k,t)]\,dk}{\int k^{2}\,M(k,t)\,dk},
\end{align}
which accounts for the contribution from the full dissipation spectrum and therefore 
provides a more reliable estimate of the dissipation scales. The resulting ratio of the 
dissipation length scales $\ell_{\nu}/\ell_{\eta} = k_{\eta}/k_{\nu}$. 

The time-averaged ratio $\langle\ell_{\nu}/\ell_{\eta}\rangle$ in the saturated phase 
of the dynamo is shown in the last column in Table~\ref{tab:lscales}, computed over 
the same number of independent realizations as done for the ratio of integral scales.
For $\Pm = 1$, we obtain $\langle\ell_{\nu}/\ell_{\eta}\rangle \sim 1.36$. 
At higher $\Pm$, the ratios increase to $\sim 2$ at $\Pm = 10$. Our results thus 
preserve the trend that the separation between $\ell_{\nu}$ and $\ell_{\eta}$ widens 
as $\Pm$ increases.

\section{Stretching versus Compression}
\label{s:energy}

Magnetic field amplification via dynamo action in supersonic turbulence results from 
a complex interplay of random stretching and compression of magnetic field lines. 
Understanding how these two processes compete and whether their relative influence 
depends on the magnetic Prandtl number ($\Pm$) is one of the key objectives of our 
work. To investigate this, we examine the PDFs of 
local stretching and compression that contribute to the growth or decay of magnetic 
energy.

Following \citet{SS24}, the evolution equation of the magnetic energy in terms of 
local stretching, advection, compression, and dissipation terms is
\begin{align}
\frac{\partial}{\partial t}\left(\frac{B^{2}}{2}\right) & = \underbrace{B_{i}B_{j}\frac{\partial U_{i}}{\partial x_{j}}}_\text{stretching} 
- \underbrace{B_{i}\frac{\partial U_{j}}{\partial x_{j}}B_{i}}_\text{compression} 
- \underbrace{B_{i}U_{j}\frac{\partial B_{i}}{\partial x_{j}}}_\text{advection} \nonumber \\
& - \underbrace{\eta B_{i}(\nabla\times\JJ)_{i}}_\text{dissipation}. 
\label{emevol_terms}
\end{align}

The velocity gradient $\partial U_{i}/\partial x_{j}$ in the above equation can be further 
decomposed as a sum of three terms: a rate of strain tensor,
\begin{align}
\label{sij}
S_{ij} = (\partial U_{i}/\partial x_{j} + \partial U_{j}/\partial x_{i})/2 - (\partial_{k}U_{k})\delta_{ij}/3, 
\end{align}
a rate of expansion tensor, $(\partial_{k}U_{k})\delta_{ij}/3$; and an antisymmetric tensor, 
\begin{align}
\Omega_{ij} = \epsilon_{ijk}\omega_{k}/2,
\label{omega}
\end{align}
corresponding to the vorticity. It is straightforward to show that the antisymmetric part 
does not contribute to the magnetic energy, as it is proportional to 
$\BB\cdot(\BB\times\omega) = 0$. Substituting equations~\ref{sij} and \ref{omega}
in \ref{emevol_terms}, we get

\begin{align}
\frac{\partial}{\partial t}\left(\frac{B^{2}}{2}\right) & = \underbrace{B_{i}B_{j}S_{ij}}_\text{stretching} 
+ \underbrace{\frac{1}{3}B_{i}B_{j}\delta_{ij}(\partial_{k}U_{k})}_\text{expansion}
- \underbrace{B_{i}\frac{\partial U_{j}}{\partial x_{j}}B_{i}}_\text{compression} \nonumber \\
& - \underbrace{B_{i}U_{j}\frac{\partial B_{i}}{\partial x_{j}}}_\text{advection}
- \underbrace{\eta B_{i}(\nabla\times\JJ)_{i}}_\text{dissipation}. 
\label{em_diffterms}
\end{align}

It is worth noting that the presence of ($\partial_{k}U_{k}$) in $S_{ij}$ implies that 
local stretching is also influenced by flow compressibility. For incompressible flows, this 
effect is negligible. The expansion and compression terms in Equation ~\ref{em_diffterms} 
combine into a term $\propto \nabla\cdot\UU$, while the advection term can be rewritten 
as $U_{j}\partial_{j}(B_{i}^{2}/2)$. Multiplying both sides of Equation ~\ref{em_diffterms} by 
$\ted/B^{2}_{\rm rms}$ we obtain the volume-integrated magnetic energy evolution 
in dimensionless form,
\begin{align}
\int_{V}\frac{\partial}{\partial t}\left(\frac{|\BB|^{2}}{2}\right)\,\frac{\ted}{B_{\rm rms}^{2}}~\dd V &=
+ \int_{V}S_{ij}B_{i}B_{j}\frac{\ted}{B_{\rm rms}^{2}}~\dd V \nonumber \\
&- \frac{2}{3}\int_{V}|{\BB}|^{2}({\nabla\cdot\UU})\frac{\ted}{B_{\rm rms}^{2}}~\dd V \nonumber \\
&- \int_{V}\UU\cdot\frac{1}{2}\nabla|\BB|^{2}\frac{\ted}{B^{2}_{\rm rms}}~\dd V \nonumber \\
&- \eta\int_{V}\BB\cdot(\nabla\times\JJ)\frac{t_{\rm ed}}{B_{\rm rms}^{2}}~\dd V.
\label{emevol_dimless}
\end{align}

Following the approach described in \citet{SS24}, we compute the PDFs for the dimensionless stretching and compression terms in Equation ~\ref{emevol_dimless}, 
and derive the corresponding mean values, $\xi_{\rm s}$ (stretching) and $\xi_{\rm c}$ (compression).
Table~\ref{tab:mpdfs_sum} lists these means averaged over several independent realizations 
of the dynamo (at different $\ted$), in the kinematic and saturated phases. 

\begin{table}[h!]
\centering
\setlength{\tabcolsep}{3pt}
\begin{tabular}{llcc|cc}
\hline
Run & $\Pm$ & \multicolumn{2}{c|}{$\xi_{\rm s}$} & \multicolumn{2}{c}{$\xi_{\rm c}$} \\
\cmidrule(lr){3-4} \cmidrule(lr){5-6}
 &  & Kin. & Sat. & Kin. & Sat. \\
\midrule
Pm1 & 1 & $7.07 \pm 0.31$ & $2.33 \pm 0.15$ & $8.35 \pm 0.23$ & $1.77 \pm 0.13$\\
Pm5 & 5 & $4.57 \pm0.21$ & $2.22 \pm 0.07$ & $4.48 \pm 0.18$ & $1.40 \pm 0.12$ \\
Pm10 & 10 & $3.59 \pm 0.16$ & $1.62 \pm 0.11$ & $2.85 \pm 0.21$ & $0.56 \pm 0.10$ \\
\hline
\end{tabular}
\caption{Time-averaged Values of $\xi_{\rm s}$ and $\xi_{\rm c}$ Obtained from 
the PDFs of Stretching and Compression Terms in Equation ~\ref{emevol_dimless} in the 
kinematic and saturated phases.}
\label{tab:mpdfs_sum}
\end{table}

In line with the expectation that both stretching and compression are influenced by flow 
compressibility, our analysis across all three simulations reveals a consistent decrease 
in both $\xi_{\rm s}$ (stretching) and $\xi_{\rm c}$ (compression) as the dynamo 
transitions from the kinematic to the nonlinear saturated phase. Interestingly, and in 
contrast to the expectation that line stretching should dominate over compression, Pm1 
shows that magnetic energy growth during the kinematic phase is actually driven primarily 
by compression, with stretching remaining subdominant by a factor of $\approx 1.2$. This 
is because both viscous and resistive dissipation act on comparable scales such that the 
shocks remain effective in driving density compressions that amplify the fields. However, 
as the dynamo evolves into the saturated phase, both mechanisms are suppressed. The 
suppression is significantly stronger for compression, with $\xi_{\rm c}$ decreasing by a 
factor of $\approx 4.7$ compared to a decrease by a factor of about 3 for $\xi_{\rm s}$. 

In Pm5, stretching and compression are initially comparable ($\xi_{\rm s} \approx \xi_{\rm c}$)
in the kinematic phase, but compression again declines more sharply in the saturated 
regime. For Pm10, the behavior shifts. Here, line stretching clearly dominates over 
compression in the kinematic phase, by a factor of about 1.26. This is due to the fact that 
a higher viscosity suppresses velocity fluctuations on scales larger than the resistive scales. 
This suppression reduces the compressive motions that are more easily dissipated by 
viscosity. As a result, line stretching from vortical motions becomes the dominant agent in 
amplifying the field. Notably, in the saturated phase, compression is suppressed even more 
dramatically with $\xi_{\rm c}$ dropping by a factor of $\sim 5$ compared to its kinematic 
value -- greater than the suppression seen for $\Pm = 1$ and $5$. In all three runs, 
the stronger suppression of compressive motions relative to stretching once again 
reinforces the role of magnetic pressure forces in resisting further compression \citep{SS24}.

\section{Faraday Rotation from $3N^{2}$ lines of sight}
\label{s:faraday}

\begin{figure}[t!]
\begin{centering}
\includegraphics[width=\columnwidth]{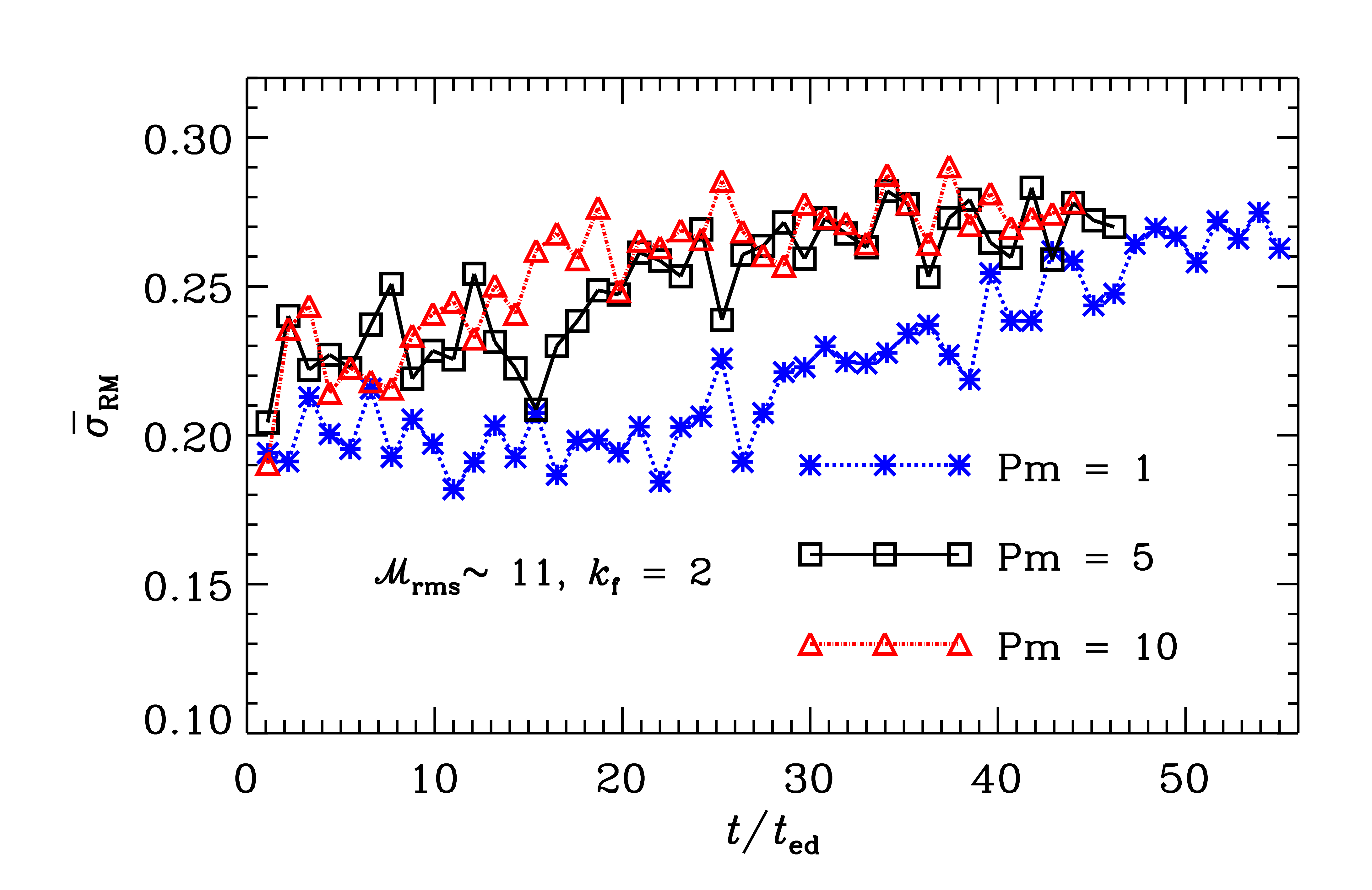} 
\end{centering}
\caption{Time evolution of $\sigmarm$ with $t/\ted$ for all runs listed in Table~\ref{tab:sumsim}.}
\label{fig:sigmarm_m11_diffpm}
\end{figure}

Faraday rotation serves as a key probe of the line-of-sight (LOS) component of 
magnetic fields in astrophysical plasmas. In a magnetized environment such as 
the galactic ISM, the polarization angle of linearly polarized 
radio emission undergoes a wavelength-dependent rotation. The angle of rotation 
increases with the square of the wavelength, and the proportionality factor is 
known as RM, 
\begin{align}
{\rm RM} = K \int_{L} n_{\rm e} \,\BB \cdot d\boldsymbol{\ell},
\end{align}
where $n_{\rm e}$ is the thermal electron density, $\BB$ is the magnetic field vector, 
and the integration is along the LOS $L$ from the source to the observer. The constant 
$K = 0.81\,{\rm rad\,m^{-2}\,cm^{3}\,\mu G^{-1}\,pc^{-1}}$ encapsulates physical 
constants. In what follows, we explore how coherent the fields generated by 
fluctuation dynamos in supersonic turbulence are and assess how this coherence varies 
with $\Pm$. Since supersonic flows result in significant density fluctuations along the 
LOS, we retain $n_{\rm e}$ inside the integral for all the runs. Following the methodology 
outlined in \citet{SBS18}, we compute $\int{\rho\BB\cdot d\bfl}$ directly for each simulation 
listed in Table~\ref{tab:sumsim}, evaluating the RM over $3N^{2}$ LOSs, 
along each of the $x, y$ and $z$-directions. For example, if the LOS integration is along 
$z$, the RM at a transverse position $(x_{i}, y_{i})$ is given by the discrete sum of $B_{z}$, 
\EQ
{\rm RM} (x_{i}, y_{i}, t) = \frac{K}{\mu m_{p}} \sum_{j=0}^{N-1} 
\left(\frac{L}{N}\right)\,\rho B_{z}\left(x_{i}, y_{i}, \frac{L}{N}j, t\right), 
\label{rmeq}
\EN
where $n_{\rm e} = \rho/\mu m_{p}$ is expressed in terms of the density, $L$ is the 
box length, $N$ is the number of grid points along each axis, $\mu = 0.61$ is the mean 
molecular weight, and $m_{p}$ is the proton mass. 
Because our simulations are isothermal, the gas temperature is fixed and 
the ionization fraction is implicitly constant. Thus, the thermal electron density 
scales linearly with the gas density ($n_{\rm e} \propto \rho$), resulting in a 
perfect point-point correlation. In a realistic ISM, however, this correlation can vary 
substantially as radiative cooling and recombination lower the ionization fraction 
in dense, cold gas, leading to a corresponding reduction in $n_{\rm e}$ \citep[e.g.,][]{Bracco+22}. 
However, the dominant contribution to the observable RM typically arises from 
the warm ionized medium (WIM) and other diffuse ionized phases \citep{H15,SF22}, 
where the electron fraction is relatively high and hence the assumption of a local 
proportionality between $n_{\rm e}$ and $\rho$ remains a reasonable approximation. 
In this sense, the RM analysis presented here applies to an idealized, WIM-like 
medium. This provides a clean and unambiguous framework for quantifying 
the magnetic coherence scale produced by the fluctuation dynamo in highly 
compressible turbulence, without the additional complications introduced by 
a multiphase ISM.

Magnetic fields generated by the fluctuation dynamo are expected to be nearly 
statistically isotropic, leading to a vanishing mean RM, i.e., $\langle \int\rho\BB\cdot d\bfl\rangle=0$. 
We thus concentrate on the standard deviation of the RM, $\sigma_{\rm RM}$, and 
examine its time evolution across different values of $\Pm$. 
To facilitate comparisons across different runs and physical regimes, we further normalise 
$\sigma_{\rm RM}$ by a characteristic value 
\EQA
\sigma_{\rm RM0} &=&  \frac{K}{\mu m_{p}} \frac{(\rho B)_{\rm rms}}{\sqrt{3}} L \sqrt{\frac{2\pi}{k_{\rm f}L}}, 
\label{sigmarm_dimless}
\ENA 
assuming that magnetic fields (weighted by density) are randomly oriented within turbulent 
cells of correlation length $l_{\rm f} = 2\pi/k_{\rm f}$ in a box of size $L$. This assumption 
is justified for supersonic flows, where density fluctuations could significantly vary from one 
turbulent cell to another and can also be correlated with magnetic field variations. 
Using the above normalization, we next compute the evolution of the normalized standard 
deviation $\sigmarm=\sigma_{\rm RM}/\sigma_{\rm RM0}$ of the set ${\rm RM}(x_{i}, y_{i}, t)$. 
For LOSs along $x$ and $y$, $B_{x}$ and $B_{y}$ values are to be used, respectively, in 
Equation ~\ref{rmeq}. The final $\sigmarm$ value is obtained by averaging the standard 
deviations computed along the three principal directions. For reference, a fluctuation-dynamo-generated field ordered on the forcing scale would have $\sigmarm \sim 1$.

Figure~\ref{fig:sigmarm_m11_diffpm} illustrates the temporal evolution of $\sigmarm$ for 
the three runs, differing in $\Pm$ values. Overall, in each simulation, $\sigmarm$ starts 
from a value of $0.2$ and then increases as the fluctuation dynamo starts to amplify the 
field. As $\sigmarm$ rises above $\sim 0.22$, subtle differences between the runs emerge. 
For Pm1 (blue asterisks), $\sigmarm$ continues to fluctuate around $\sim 0.2$ until 
$t/\ted \approx 23$ after which it increases slowly until $t/\ted \approx 46$. Even though 
the magnetic energy saturates by $t/\ted \approx 40$ (see Fig.~\ref{fig:tseries_mrms_emag}), 
the slow increase of $\sigmarm$ reflects the fact that Lorentz forces continue to reorganize 
the field, which may modestly increase the magnetic coherence scale. This structural 
adjustment naturally produces a residual evolution in $\sigmarm$ up to 
$t/\ted \approx 46$. Additionally, since $\sigmarm$ is an integral quantity, it is sensitive 
to intermittency and variations in ($\rho - B$) correlations across many sight lines. 
Therefore, snapshot-to-snapshot variations are expected and are indeed seen in 
each of the three curves in the figure. On the other hand, efficient dynamo action in the 
$\Pm > 1$ runs leads to a gradual increase in $\sigmarm$ for Pm5 (black squares) 
and Pm10 (red triangles), noticeable after $t/\ted \approx 10$. This early increase 
in $\sigmarm$ could be attributed to the fact that at $\Pm > 1$, eddies richer in 
vorticity are less impeded by viscosity. Beyond $t/\ted \approx 20$, the curves for 
$\Pm = 5$ and $10$ settle into a quasi-steady state, while the curve for $\Pm = 1$ 
keeps evolving and only reaches a steady state at $t/\ted \approx 44$. 
Despite the difference in evolution, we find that the time-averaged 
$\langle \sigmarm \rangle \approx 0.26\pm 0.005$ for Pm1, $\approx 0.26 \pm 0.01$ 
for Pm5, and $\approx 0.27 \pm 0.009$ for Pm10. These values are computed 
using nine realizations for Pm1 and 17 realizations each for Pm5 and Pm10 over the
intervals $t/\ted = 47-55, 25-46$, and $25-45$, respectively. This implies that the 
steady state value of $\sigmarm$ is nearly independent of $\Pm$ once the dynamo 
saturates but is lower for $\Pm = 1$ in the kinematic phase. The similarity in 
the steady-state values of $\langle \sigmarm \rangle$ across all the runs imply that 
the effect of the Lorentz forces is to order the field on a maximum scale that 
depends only on the forcing scale. In our simulations, this Lorentz-force-regulated 
magnetic coherence scale saturates at $\approx (1/4 - 1/3)\lf$, when the driving 
scale is at $1/2$ the scale of the simulation domain.

\section{Conclusions}
\label{s:conc}

Fluctuation dynamos provide a key mechanism for amplification of seed magnetic 
fields in random/turbulent flows by the action of turbulent eddies. In the ISM which 
is both highly conducting and compressible, these dynamos operate by a combination 
of random (in time) stretching and compression of magnetic field lines 
\citep[e.g.,][]{Fed+11,Fed16,SF21a,SS24}. In this work, we have carried out a 
systematic study of fluctuation dynamos in supersonic turbulence ($\mrms \approx 11$), 
investigating their dependence on the magnetic Prandtl number up to $\Pm = 10$. 
In what follows, we delineate the important findings from our work with an emphasis
on the distinct characteristics obtained between $\Pm = 1$ and $\Pm > 1$ regimes. 

The time evolution of $E_{\rm m}/E_{\rm k}$ shows a clear dependence on $\Pm$ 
with larger $\Pm$ yielding faster growth rates and higher saturation levels than 
$\Pm = 1$. The 2D slices of $\log(\rho/\langle \rho\rangle)$ display 
the characteristic features of supersonic turbulence, namely, sharp density contrasts 
where dense structures are surrounded by under dense voids. We also find the 
in-plane magnetic fields to be arranged in folds and strongly compressed at myriad 
locations indicating field amplification by a combination of local stretching 
and compression. Differences between the $\Pm = 1$ and $\Pm > 1$ regimes also 
emerge in the time evolution of the Pearson correlation coefficient '$r_{p}$' of the 
density and magnetic field strength. Specifically, during the transition from the kinematic 
to the nonlinear saturated phase, $r_{p}$ decreases only by a factor of 1.2 for 
$\Pm = 1$ but by $\sim 2$ for $\Pm = 10$. This stronger suppression for $\Pm = 10$ 
reflects the role of magnetic pressure forces in resisting the continued compression 
of the field lines due to supersonic turbulence. In fact, as Fig.~\ref{fig:align} clearly 
illustrates, the antiparallel alignment between $\nabla \rho$ and $\nabla B^{2}$ is 
more pronounced for $\Pm = 10$ than for $\Pm = 1$. Moreover, separating regions 
by field strength reveals that the enhanced suppression of $r_{p}$ for $\Pm = 10$ 
arises mainly from the reduced correlation between $\rho$ and $B$ in sub-rms 
($B/\brms \leq 1$) regions, although the correlation also weakens in the strong field 
($B/\brms > 1$) regime.

Analysis of the solenoidal and compressive kinetic energy spectra shows that,
for $\Pm = 10$, higher viscosity suppresses small-scale velocity fluctuations 
more strongly in compressive than in solenoidal modes. Consequently, the 
solenoidal-to-compressive kinetic energy ratio is larger than in the $\Pm = 1$ 
case. Estimates of the integral scales of the velocity and magnetic fields in 
the saturated phase show that the ratio of $\ell^{V}_{\rm int}/\ell^{M}_{\rm int}$ 
is independent of $\Pm$, yielding values of $\sim 3.4$. This suggests that even 
in the nonlinear phase, the magnetic energy remains peaked on scales $k > \kf$. 
On the other hand, our numerical estimates of the viscous-to-resistive 
dissipation scale ratio exhibit an increase with $\Pm$, with 
$\langle\ell_{\nu}/\ell_{\eta}\rangle \approx 1.36$ for $\Pm = 1$, 
$\approx 1.70$ for $\Pm = 5$, and $\approx 1.92$ for $\Pm = 10$. 
This monotonic behavior is consistent with the expectation that higher $\Pm$ 
yields a broader separation between viscous and resistive scales.

Comparison of the contributions from line stretching and compression reveals 
key differences between $\Pm = 1$ and $\Pm > 1$. Note that in incompressible 
flows, the decline in $\xi_{\rm s}$ during the nonlinear phase reflects reduced 
stretching \citep{SS24}. In compressible flows, however, both stretching and 
compression are modified by density fluctuations. For $\Pm = 1$, we find 
$\xi_{\rm c} > \xi_{\rm s}$ in the kinematic phase, indicating that field amplification 
is initially driven by compression associated with density enhancements. As the 
system saturates, both $\xi_{\rm s}$ and $\xi_{\rm c}$ decrease, with stronger 
suppression of $\xi_{\rm c}$. For $\Pm = 5$, $\xi_{\rm s} \approx \xi_{\rm c}$ in 
the kinematic phase, but stretching again dominates in the saturated phase. 
However, for $\Pm = 10$, amplification in the kinematic phase is initially driven 
by random stretching rather than compression, likely due to the higher viscosity 
($\Rey \approx 660$) which hinders flux tube compression. In the saturated 
phase, both terms decrease, but $\xi_{\rm c}$ is suppressed by nearly a factor 
of 5. Irrespective of the $\Pm$, the stronger suppression of $\xi_{\rm c}$ in the 
saturated phase reflects the role of magnetic pressure opposing further 
compression in agreement with \citep{SS24}. 

In summary, the aforementioned results in the saturated phase show
the following trends : (i) $\langle r_{p}\rangle$ decreases with increasing 
$\Pm$; (ii) $\langle\ell^{V}_{\rm int}/\ell^{M}_{\rm int}\rangle$ remains 
constant across $\Pm$, while $\langle \ell_{\nu}/\ell_{\eta}\rangle$ increases 
with $\Pm$; and (iii) the PDFs of stretching and compression terms reveal 
that in the kinematic phase, compression dominates for Pm1, both terms 
are comparable for Pm5, and stretching dominates for Pm10. In all three 
runs, stretching dominates over compression in the saturated phase. 
To assess the statistical significance of these results, we use the mean ($\mu$) 
and the $1\sigma$ scatter of the relevant variables together with the 
number of realizations '$N$' to compute the standard error of the mean, 
${\rm SEM} = \sigma/\sqrt{N}$. Differences between runs are then 
evaluated via pairwise comparisons of the mean values, taking into account 
the combined SEMs of each pair of runs\footnote{For two runs '$A$' and '$B$' 
with means $\mu_{A}$ and $\mu_{B}$, the pairwise difference is 
$\Delta = |\mu_{A} - \mu_{B}|$, and the combined SEM is 
${\rm SEM}_{c} = \sqrt{{\rm SEM}^{2}_{A} 
+{\rm SEM}^{2}_{B}}$.} \citep{BR03,WJ12}.

The SEMs of $\langle r_{p}\rangle$ for Pm1, Pm5, and Pm10 are 
$7\times 10^{-3}, 3\times 10^{-3},$ and $2.89\times 10^{-3}$, respectively.
The pairwise differences ($0.08-0.19$) exceed the combined SEMs by 
factors of ($10-25$), confirming that the decline in $\langle r_{p}\rangle$ 
with increasing $\Pm$ is statistically significant. In contrast, pairwise 
differences for $\langle\ell^{V}_{\rm int}/\ell^{M}_{\rm int}\rangle$
are $\leq 0.03$, smaller than the corresponding SEMs, consistent with 
the lack of significant $\Pm$ dependence. Conversely, for 
$\langle\ell_{\nu}/\ell_{\eta}\rangle$, the mean differences ($0.22-0.56$) 
are $1-2$ orders of magnitude larger than the SEMs, indicating that 
the increase in $\langle\ell_{\nu}/\ell_{\eta}\rangle$ with $\Pm$ 
is statistically robust. Similar analyses for the stretching and compression 
terms show that at Pm1, the strong compression dominance in the 
kinematic phase is confirmed by the pairwise difference being $\sim 12$ 
times larger than the combined SEM. For Pm5, stretching and compression 
become statistically indistinguishable, with the difference only $\sim1.2$ 
times the combined SEM, indicating a transitional regime where neither 
mechanism dominates. At Pm10, stretching already dominates in the 
kinematic phase, with the difference exceeding the combined SEM by a 
factor of $\sim10$. In all three runs, the saturated phase shows a robust 
dominance of stretching over compression, with differences ($8-25$) times 
larger than the combined SEMs, reflecting the suppression of compressive 
amplification once the Lorentz force becomes dynamically important.

Finally, we examined the coherence of fluctuation-dynamo-generated 
fields in supersonic turbulence ($\mrms \approx 11$). Results from our 
synthetic RM measurements for $\Pm = 1-10$ yield coherence scales 
$\ell_{\rm c} \sim (1/4 - 1/3)\lf$ for driving at half the box scale. 
Using the mean and $1\sigma$ values from Section~\ref{s:faraday} and 
$N = 9, 17, 17$ we obtain SEMs ($1.6\times 10^{-3}, 2.4\times 10^{-3}, 
2.1\times 10^{-3}$), for Pm1, Pm5 and Pm10, respectively. Analysis of 
the difference between the means suggests that while Pm1 and Pm5 
are statistically indistinguishable, Pm10 shows only a very small difference 
($\sim 3.6\%$). This confirms that within the range of $\Pm$ values 
explored in this work, $\sigmarm$ is nearly independent of $\Pm$.

In recent years, Mg II absorption systems studied by 
\citet{Bernet+08, BML10, Farnes+14} and \citet{MCS20} have revealed excess 
Faraday rotation consistent with ordered magnetic fields of $\mkG$ strength 
in galaxies out to $z\sim 1$, when the Universe was only 6 billion years old. 
This motivates assessing whether fluctuation dynamos in compressible 
turbulence can account for a significant component of the RM signal. 

For a physical estimate, we note that the total stellar mass of the Milky Way 
is $\sim 2.6\times 10^{10}M_{\odot}$ \citep{Lian+25}, with a gas mass fraction 
assumed to be of order $10\%$. Assuming that this gas is distributed in a disk 
of radius $r = 10\kpc$ and disk thickness $2h = 1\kpc$, the average number 
density $n \sim 0.4\cmn$. It is plausible that high-redshift star-forming disks, 
may exhibit substantially larger gas fractions, so the corresponding electron 
densities may exceed those typical of the Milky Way WIM. Indeed, nebular 
diagnostics of [S II] and [O II] doublets in $z \sim 1-2$ galaxies yield internal 
HII region electron densities, $n_{\rm e} \approx (50 - 300)\cmn$ \citep{K+17, 
Davies+21}. If such ionized clumps have modest filling factors 
$f_{V} \sim 0.01 - 0.1$, the resulting $\edense \sim (0.5 - 30)\cmn$. 
For simplicity, and to remain conservative, we adopt $\edense \sim 1\cmn$. 

Assuming typical vortical turbulent velocities of $\urms \sim 10\kms$ of the 
order of the sound speed in the WIM and forcing scale of $\lf \sim 100\pc$, 
the eddy turnover time is $\ted \sim 10^{7}\yr$, enabling the fluctuation 
dynamo to grow and saturate the fields well within the lifetime of disk galaxies. 
For a path length of $L = 1\kpc$ through the disk thickness, the normalization 
factor is
\EQA
\sigma_{\rm RM0} &\sim& 444\,\radm\,\left(\frac{\overline{n}_{\rm e}}{1\, {\rm cm^{-3}}}\right) \nonumber \\
&&\times \left(\frac{B_{\rm rms}}{3 \mkG}\right)\,
\left(\frac{L}{1\,{\rm kpc}}\right)^{1/2}\,\left(\frac{l_{\rm f}}{100\,{\rm pc}}\right)^{1/2},
\label{phys_sigmarm}
\ENA
obtained from a simple model of random magnetic fields, where the fields 
are assumed to be random with a correlation length $\lf = 2\pi/\kf$ 
\citep[e.g.,][]{Soko+98, SSH06, CR09, BS13,SBS18,SF21b}. 
The equipartition field $B_{\rm eq} = (4\pi\,\rho\,u^{2}_{\rm rms})^{1/2} 
\sim 5\mkG$. If the fluctuation dynamo saturates at a fraction $f \sim (0.1-0.2)$ 
of equipartition, we obtain $\brms = f\,B_{\rm eq} \approx (0.5-1)\mkG$\footnote{The 
magnetic field in clouds denser than the average ISM could be larger than 
the $\brms$ estimated above.}, giving $\sigma_{\rm RM0} \sim (74-150)\radm$. 
Thus, for $\sigmarm \sim 0.3$, the Faraday RM dispersion is $\sim (22-45)\radm$. 
This level of RM dispersion is comparable to the RM excess values reported 
for Mg II absorbers, which range from tens of $\radm$ \citep{Farnes+14} to 
$\geq 100\radm$ for strong absorbers \citep{Bernet+08}. This suggests that 
fluctuation dynamos in gas-rich, turbulent disks at $z\sim1$ could plausibly 
account for a substantial fraction of the observed RM signal. However, 
several caveats apply: (i) our estimates assume a uniform disk geometry 
rather than clumpy circumgalactic medium/ISM structures; (ii) the true filling factors, thermal 
electron densities, efficiency of the dynamo, and turbulent driving scales in 
high-$z$ galaxies are uncertain; and (iii) Mg II absorbers may probe both 
disk and halo gas, whereas our estimates pertain strictly to disklike 
environments. In addition, observed extragalactic RMs include contributions 
from the Galactic foreground, intrinsic source rotation, and redshift-dependent 
depolarization, all of which introduce systematic uncertainties when attributing 
the RM signal to intervening galaxies. In this context, our estimates of RM 
dispersions that are broadly comparable to observationally inferred RM 
excesses offer a useful theoretical framework for their interpretation.

We further emphasize that in supersonic turbulence, the use of a 
volume-averaged $\edense$ in estimating $\sigma_{\rm RM0}$ is an 
approximation, valid primarily for order-of-magnitude comparisons. 
In practice and as shown earlier, density fluctuations can be significant 
and correlated with magnetic field variations. Our simulation-based RM 
values retain the full spatial variation of $\rho$ within the LOS integrals to 
capture these effects accurately.

In a realistic ISM, the ionized phases typically occupy only a fraction 
$f_{V} < 1$ of the total volume. Let us consider an LOS of 
total length '$L$' that passes through '$N$' turbulent cells of size $\lf$. 
If only a fraction $f_{V}$ of the LOS is filled with the ionized gas, then 
$N = f_{V}L/\lf$, instead of $N = L/\lf$ for $f_{V} = 1$. Consequently, 
$\sigma_{\rm RM} \propto f_{V}^{1/2}$, whereas the normalized quantity 
$\sigmarm$ is independent of $f_{V}$. Thus, although the absolute RM 
dispersion decreases by $f_{V}^{1/2}$, the normalized RM statistics 
reported in our work remain representative of the intrinsic magnetic 
coherence properties.

Given that the ISM is inherently multiphase, a natural next step is to 
extend this work to more realistic simulations that include cooling, 
heating, and phase structure. Such studies would clarify how spatially 
varying $\Pm$ and different driving mechanisms shape the fluctuation 
dynamo and how these effects propagate into the resulting 
magnetic field topology and Faraday RMs.

\section*{acknowledgments}
A.N. and S.S. acknowledge the use of the High 
Performance Computing resources made available by the Computer 
Centre of IIA. S. S. and B. V. thank Mr.~Sayeed Kazim H. 
for testing an earlier version of the turbulence driving module as part of 
his MSc thesis project at IIT Indore. The authors thank Prof.
Andrea Mignone for valuable discussions and the reviewer for
a timely and constructive report.

\software{Astropy~\citep{astropy}, h5py~\citep{hdf5}, Jupyter~\citep{jupyter},  Matplotlib~\citep{matplotlib}, 
Numpy~\citep{numpy_nature}, Scipy~\citep{scipy}.}

\appendix

\section{Alignment Angles}
\label{sec:align}

\begin{figure}[t!]
\begin{centering}
\includegraphics[width=\columnwidth]{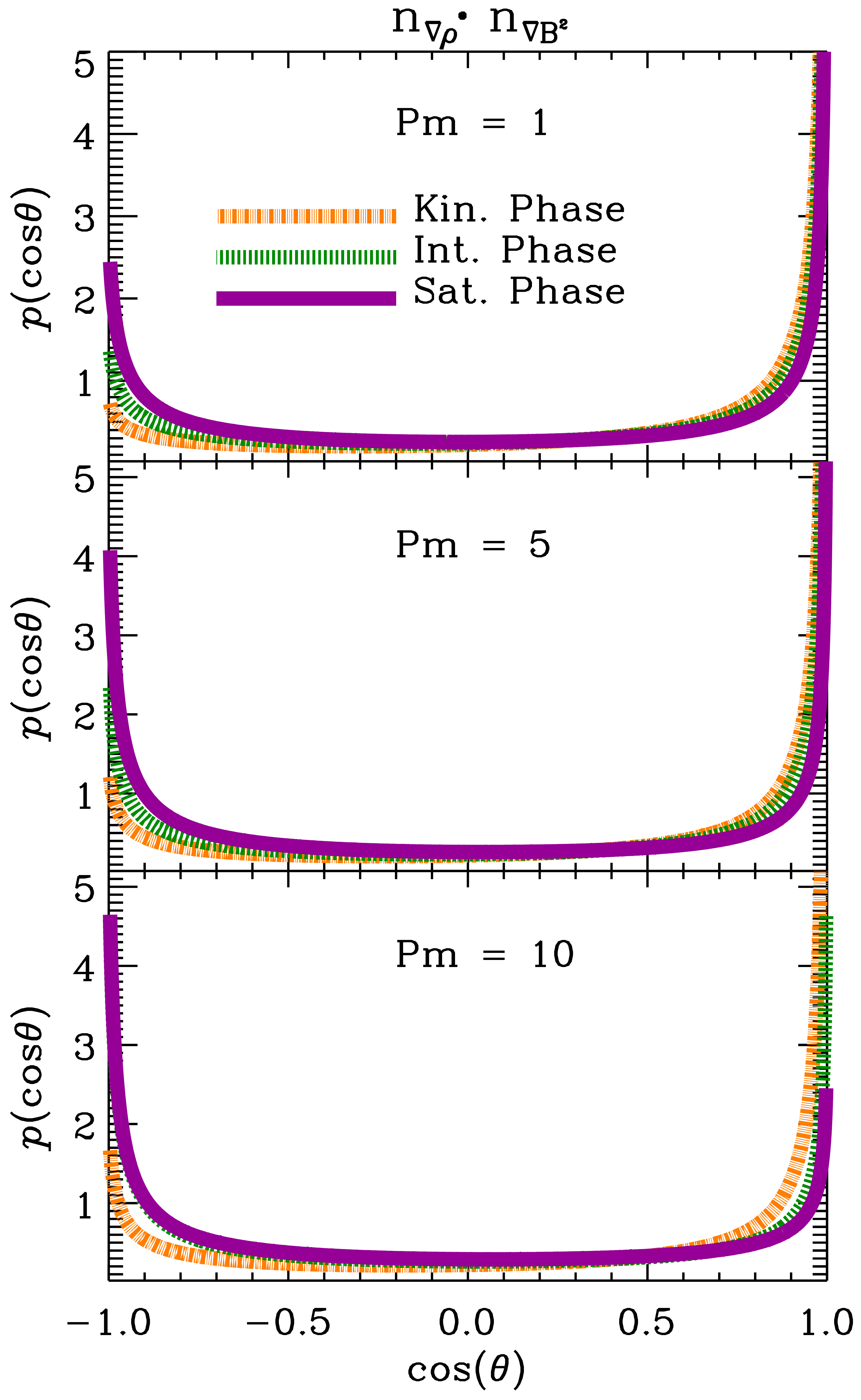} 
\end{centering}
\caption{PDFs of the cosine of the angle between $n_{\nabla\rho}$ and $n_{\nabla B^{2}}$ 
for runs Pm1 (top), Pm5 (middle), and Pm10 (bottom). For each $\Pm$, the distributions are 
computed during the kinematic (orange), intermediate (green), and saturated 
(purple) phases of fluctuation dynamo evolution using multiple independent realizations.}
\label{fig:align}
\end{figure}

In Figure~\ref{fig:align}, we present the PDFs of the 
cosine of the angle between the gradient of the density and the gradient of magnetic 
pressure for Pm1 (top), Pm5 (middle), and Pm10 (bottom). For each case, the PDFs 
are shown during the kinematic (orange), intermediate (green), and saturated (purple) 
phases of magnetic field evolution. These PDFs are computed over multiple independent 
realizations (at different $\ted$) of the fluctuation dynamo in each of the aforementioned 
phases. As evident from the figure, during the kinematic phase, density enhancements 
play a crucial role in driving the amplification of the magnetic field. Consequently, the 
strong positive correlation observed in Fig.~\ref{fig:corr_m11_diffpm} indicates that 
density variations are also correlated with variations in magnetic pressure. This results 
in an initial parallel alignment between $\nabla\rho$ and $\nabla B^{2}$. This trend 
can be inferred from the PDF values at $\cos(\theta) = 1$. For both Pm1 and Pm5, 
the density at $\cos(\theta) = 1$ remains close to $\sim 5$ across all three phases.
However, as the magnetic field continues to be amplified through random stretching, 
magnetic pressure forces begin to resist further compression of the field lines, leading 
to the emergence of an antiparallel alignment between the two, consistent with 
the findings of \citet{SS24}. This can again be seen from the density at 
$\cos(\theta) = -1$, which increases to $\sim 2.5$ from its kinematic value.

However, we observe a subtle dependence on $\Pm$, particularly in the saturated 
phase. As $\Pm$ increases, the antiparallel alignment becomes more pronounced, 
while the parallel alignment gradually diminishes. This behavior is already 
pronounced for Pm5, where the density at $\cos(\theta) = -1$ rises to $\sim 4$ in 
the saturated phase. For Pm10, the effect is the strongest. In the kinematic phase, 
the PDF is strongly peaked at $\cos(\theta) = +1\,(p\simeq 5)$, with a much smaller 
density at $\cos(\theta) = -1\,(p\simeq 1.5)$, indicating that parallel alignment dominates. 
In the saturated phase, however, this pattern reverses. The density at 
$\cos(\theta) = +1$ drops to $\sim 2.5$, while the same at $\cos(\theta) = -1$ 
rises to $\sim 4.5$. This inversion of the relative endpoint values demonstrates a 
clear shift from strongly parallel to strongly antiparallel configurations, showing 
that antiparallel alignment becomes increasingly probable as the dynamo saturates.

Taken together, these results suggest that in the saturated phase, magnetic 
pressure forces counteract compressive motions more effectively for $\Pm > 1$ 
than for $\Pm = 1$. For $\Pm = 10$, this effect is significant enough to lead to a 
steep decline in $r_{p}$, as seen in Fig.~\ref{fig:corr_m11_diffpm}.


\end{document}